\documentclass[aps,prl,byrevtex,amsmath,amssymb,twocolumn]{revtex4}
\usepackage{epsfig}
\newcommand{\be}{\begin{equation}}
\newcommand{\ee}{\end{equation}}
\newcommand{\bea}{\begin{eqnarray}}
\newcommand{\eea}{\end{eqnarray}}

\newcommand{\D}{\Delta}

\newcommand{\lf}{\left}
\newcommand{\rg}{\right}
\newcommand{\nn}{\nonumber}

\begin{document}
\title{Exact exchange-correlation potential of a ionic Hubbard model 
   with a free surface.}

\author{V. Brosco, Z.-J. Ying, J. Lorenzana}
\affiliation{ISC-CNR and Dipartimento di Fisica, University of Rome 
``La Sapienza'', P.le A. Moro 2, I-00185 Rome, Italy }

\date{\today}
\begin{abstract}
We use Lanczos exact  diagonalization to compute the exact exchange correlation
 potential ($v_{xc}$) of a Hubbard chain with large binding energy (``the
 bulk'') followed by a chain with zero binding energy  (``the
 vacuum'').
Several results of density functional theory in the
 continuum (sometimes controversial) are verified in the  lattice. 
In particular we show explicitly that the fundamental gap is given by
the gap  in the Kohn-Sham 
 spectrum plus $\Delta_{xc}$, the jump on  $v_{xc}$ in the bulk when a
 particle is added.
The presence of a staggered potential and a nearest-neighbor
interaction $V$ 
allows to simulate a ionic solid. We show that in the small hopping 
amplitude limit $\Delta_{xc}=V$ in the ionic regime, while in the Mott regime $\Delta_{xc}$ is
determined by the Hubbard $U$ interaction. In
addition we show that correlations generates a new potential barrier in
 $v_{xc}$ at the surface. 
\end{abstract}

\maketitle

Density functional theory\cite{hohenberg,kohn,kohn1999}  (DFT)  plays a major role in our
understanding of ground state properties of materials. However
most approximate DFT approaches fail to predict 
the fundamental gap $\Delta_C$ of insulators    and semiconductors
(band gap problem)
\cite{perdew1982,perdew1983,sham1983,gunnarsson1986,godby1986,cohen2008,gruning2006,morisanchez2008,lany2008,aulbur2000},
in systems ranging from   
bulk Silicon \cite{godby1986} to ZnO \cite{lany2008} and
other correlated insulators\cite{aulbur2000}.

 At the heart of almost all practical computational schemes based on
 density functional 
 theory \cite{hohenberg,kohn,kohn1999} (DFT) lies the assumption,  
first introduced by Kohn and Sham\cite{kohn}, that the ground-state
density $\rho$ of an interacting electron gas  in an external potential can
be reproduced by a  system of non-interacting electrons in an
effective potential $v_{\rm KS}$. The effective potential  
can be expressed as the sum of  three
contributions:    the
external potential, $v$, the Hartree potential
$v_{H}$, and a term which accounts for exchange and
correlation effects, $v_{\rm xc}$.
   The latter is the functional derivative of a universal ``divine 
functional''\cite{mattsson2002} of the density whose precise form is not known. 
%
As first discussed by Perdew { \sl et al.} \cite{perdew1982,perdew1983} and by Sham and Sch\"ulter  \cite{sham1983}, the exchange-correlation potential $v_{xc}$ may
 have a jump  of order one when one particle is added to a solid. This
 jump, which is  absent in local and semi-local approximate
 functionals,\cite{cohen2008,lany2008} may account for the error on
 the fundamental gap according to \cite{perdew1983,sham1983}  
\be\label{deltac-app0}
\Delta_C=\Delta_{KS}+\Delta_{xc}.
\ee
where $\Delta_{KS}$ denotes the single-particle gap in the Kohn-Sham
non-interacting system. 
The size of this effect has been however long debated 
\cite{gunnarsson1986, godby1986,sham1988,zahariev2004,sagvolden2008, gruning2006,cohen2008,morisanchez2008}.



In
a pioneering work Gunnarsson and Sch\"{o}nhammer\cite{gunnarsson1986} studied a model of a  one
dimensional spinless  insulator and found that  $\Delta_{xc}$ is small
in the band insulating regime.  
Other authors have, however,
argued that the discontinuity should be large and it should  account for a large
part of the band gap problem\cite{godby1986,gruning2006,sham1988}. 
The elusiveness of $\Delta_{xc}$ is such that even its existence
has been recently questioned\cite{zahariev2004}.

Eq.~\eqref{deltac-app0} is based on the  DFT version of  Koopmans
theorem\cite{perdew1982,almbladh1985}   
which identifies  the ionization energy with the highest occupied Kohn-Sham
eigenvalue and whose validity has also been  subject of
controversies\cite{perdew1982,russier1992,almbladh1985,kleinman1997,perdew1997,dabo2010}.
 This debate along with the need to understand and correct the
 deficiencies of approximate DFT  approaches  has recently revived 
 the interest in small systems (zero dimensional) whose exchange-correlation
 potential can be  calculated exactly or very
 accurately\cite{damico1999,baerends2001,gorigiorgi2009,makmal2011}  or
 lattice systems where DFT or approximate lattice DFT schemes can be tested
 and analyzed in a controlled environment retaining many of the
 subtleties of the many-body problem in extended
 systems which can not be accessed otherwise\cite{lima2002,lima2003,bergfield2012,stoudenmire2011,kurth2010,stefanucci2011,evers2011,troster2012,verdozzi2011}.

\begin{figure}
\begin{center}
\includegraphics[width=0.4\textwidth]{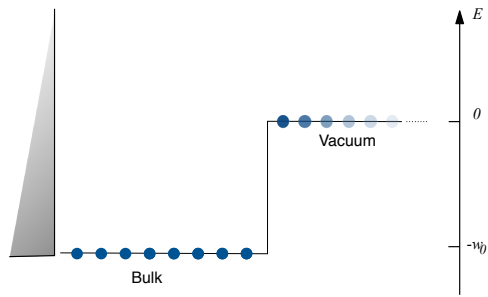}
\end{center}
\caption{Structure of the system consisting of $L_B$ bulk and $L_V$ vacuum
  sites. The bulk sites have a large binding energy,
  $w_0$. } \label{fig-system} 
 \end{figure}
%
 

 In this work we use lattice DFT to investigate the band-gap problem.
 We calculate numerically the exact exchange-correlation
potential of a correlated insulator described by a generalized Hubbard
model which can be tuned continuously from a ionic
to a Mott insulating regime\cite{nagaosa1986}. We consider an open
system with a free surface which removes any possible ambiguity
related to the validity of Koopmans theorem and or
Eq.~\ref{deltac-app0}. {\em i.e.} 
we compute each term on the left and right of  Eq.~\ref{deltac-app0}
separately which serves as a numerical test of the equation itself. 
 We find that the contribution of the exchange-correlation 
potential discontinuity to the charge gap is non-negligible 
in both regimes. 
 The presence of the surface also allows us to highlight 
the appearance of an anomaly 
in the exact exchange correlation potential in the vacuum sites which
appears as the system enters the Mott phase.

We consider a  Hubbard chain of $L_B$  sites with a large binding energy called
``the bulk'' followed by a chain of $L_V$  sites with zero binding energy termed ``the
vacuum'' with open boundary conditions as shown in
Fig.~\ref{fig-system}.  The bulk is thus a truly open system which is
crucial to completely determine the exchange-correlation potential. 

The total Hamiltonian  can be written as $H=T+H_U+H_v$ with
\bea
\label{hubb}
T&=&-t\sum_{x\sigma }(c_{x\sigma }^{\dagger}c_{x+1\sigma
} - n_{x\sigma } +H.c.)
\nn \\
H_U &=&U\sum_xn_{x\uparrow }n_{x\downarrow }+V\sum_{x\sigma} n_{x\sigma }n_{x+1\sigma }\\
H_v&=&\sum_{x\sigma  }v_{x} n_{x\sigma }\nn\label{model},\nn
\eea
where $c_{x\sigma }^{\dagger}$ creates an electron with spin $\sigma
=\uparrow ,\downarrow $ at site $x$, $U$ and $V$ are respectively the
Hubbard interaction and nearest-neighbor interaction, $t$ is the
nearest-neighbor hopping and we set $n_{x\sigma 
}=c_{x\sigma }^{\dagger}c_{x\sigma }$. 
We included a constant energy shift in the lattice kinetic energy $T$
so that single particle energies are 
measured from the bottom of the band. 
In order to simulate the work function of a solid the potential in the
bulk is taken as $v_x=-w_0 +  \delta (-1)^x $ where  $w_0 $ is a large
positive constant such that all particles in the system are bound in
the bulk region and the second term is a site dependent
potential. The potential in the vacuum is by definition $v_x=0$.


\begin{figure}[!t]
\begin{center}
\includegraphics[width=0.45\textwidth]{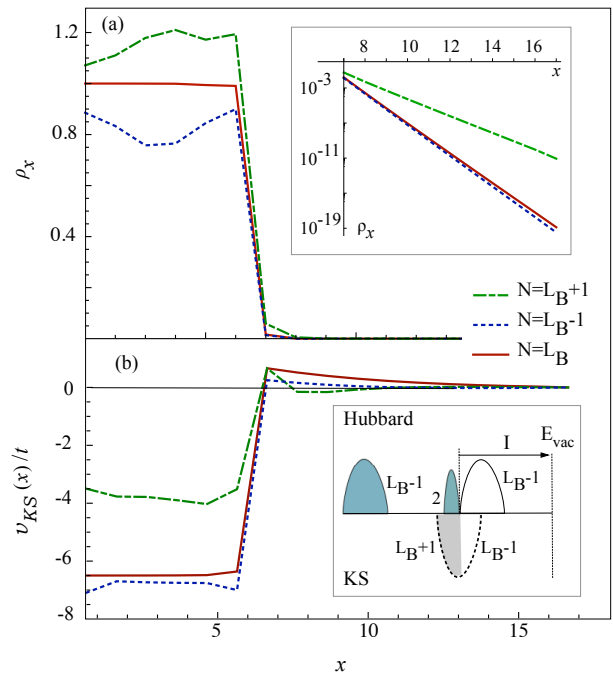}
\end{center}
\caption{ Panel (a) and (b) show respectively the charge density and the KS potential for $U=6$ at three different fillings, namely $N=L_B$, $N=L_B\pm1$. 
The inset of panel (a) presents a logarithmic plot of the density in the vacuum while the inset of panel (b) 
presents a schematic comparison between the spectrum of the Hubbard model and of the effective KS system for $N=L_B+1$. 
Other parameters: $w_0=8 t$,  $v_x=0$, $L_B=6$ and  $L_V=11$.} \label{fig-pop}
\end{figure}

We apply  DFT to this problem by considering the site occupancy
$\rho_x=\sum_\sigma \langle n_{x\sigma }\rangle $ as the fundamental
variable\cite{gunnarsson1986}. 
The charge density and the ground state energy are  obtained using
Lanczos exact diagonalization\cite{ALPS}. The exchange correlation potential is
obtained form the exact density inverting the Kohn-Sham
problem\cite{sm}.


 
In order to illustrate our capability obtain the absolute value of the
exchange correlation potential we first consider  the case of a constant external potential in the bulk  ($\delta=0$) and vanishing nearest-neighbor interaction $V=0$ . 
This corresponds  to the case of a uniform Hubbard model
which has been discussed in 
Refs.~\cite{gunnarsson1986,lima2002}. 

In the upper and lower  panels of Fig.~\ref{fig-pop}  we plot
respectively the electron density and the exact effective potential
for $U=6$ and $w_0=8$. We consider in particular the case when the
bulk is half-filled, {\em i.e.} $N=L_B$,  and   
the cases of a bulk above and below half-filling, $N=L_B\pm1$. 

 As shown in the lower panel, while the change in the potential on
the bulk on going from $N=L_B-1$ case to the $N=L_B$ is small and can
be attributed to a O($1/N$) effect, there is a sizable [O(1)] jump on 
going from  $N=L_B$ to  $N=L_B+1$. 
For all other fillings different from $N=L_B$ we find that the
 one particle addition jump is [O($1/N$)]. 

The jump for  $N=L_B$ determines 
$\Delta_{xc}$ which  we define as 
\be\label{deltaxc}
\Delta_{xc}=\sum_{x}|\varphi^{N+1}_{N+1}(x)|^2\lf(v_{\rm KS}^{N+1}(x)-v_{\rm KS}^{N}(x)\rg)
\ee
with $v_{\rm KS}^{N}$ the $N$-particle Kohn-Sham potential and 
$\varphi^{N}_{\nu}(x)$, $\epsilon_\nu^N$ the corresponding $\nu$-th
 eigenvector and eigenvalue respectively.  Notice that the shift of
 the potential is not perfectly constant in all the bulk region due to
 finite size effects. Assuming that $\varphi^{N+1}_{N+1}(x)$ is bound
 in the bulk region  Eq.~\eqref{deltaxc}
 correctly converges to the expected constant shift of the bulk in the
 thermodynamic limit. For finite systems we show below that 
 with the present definition Eq.~\eqref{deltac-app0} is satisfied with
 surprisingly small finite size corrections. Similar results 
 are obtained if the slightly different definition of
 Ref.~\cite{sham1983} is used\cite{notedeltaxc}.



\begin{figure}[!t]
\vspace{1cm}
\begin{center}
\includegraphics[width=0.45\textwidth]{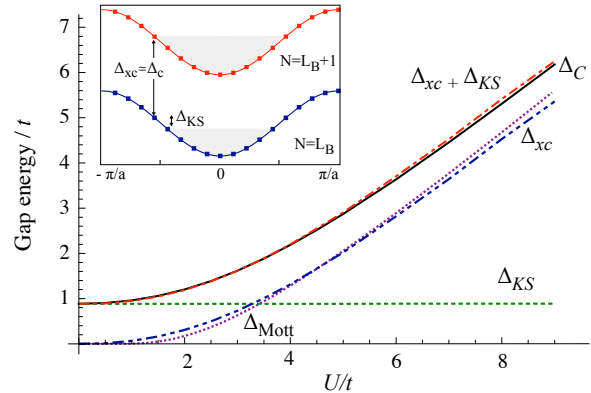}
\end{center}
\caption{Exact charge gap  $\Delta_{C}$,  Kohn-Sham gap, $\Delta_{KS}$ and contribution of the $xc$-potential jump, $\Delta_{xc}$ for a half-filled
 Hubbard chain with $L_B=6$ sites.  $\Delta_{\rm Mott}$ is the Mott
 gap for an infinite system calculated using Bethe Ansatz\cite{lieb1968}. 
The inset shows Kohn-Sham band structure of a uniform Hubbard
chain at half-filling ($N=L_B$) and with one added electron
($N=L_B+1$).
} \label{fig-gaphub}
\end{figure}

Fig.~\ref{fig-gaphub} shows the $U$ dependence of the exact charge gap of the $N=L_B$ electron
system defined as  $\Delta_C\equiv I^N-A^N$ where $I^N$ and $A^N$ indicate respectively the ionization energy and the electron affinity of the $N$-particle system,
 $I^N\equiv E^{N-1}_0-E^{N}_0$ 
 $A^N=E^{N}_0-E^{N+1}_0$ with $E^N_0$ denoting
the ground state energy of the $N$-particle system obtained with the
same Lanczos computation.   
We also show $\Delta_{xc}+\Delta_{KS}$  where $\Delta_{KS}$ is the
exact KS gap, {\sl i.e.} the gap in the spectrum of the effective
non-interacting $N$-particle Kohn-Sham system.  
%
We see that indeed 
Eq.~\ref{deltac-app0} is well fulfilled.
As discussed below the Kohn-Sham gap should vanish in the thermodynamic  
limit for a Hubbard chain so its finiteness is a finite size effect.


The charge density in the vacuum 
remains for all fillings much smaller than 1 and decays exponentially
as shown by the logarithmic plot in the inset of Fig.~\ref{fig-pop} (upper panel). 
The  change  in the density decay rate in the vacuum as the filling
becomes larger than one ($N>L_B$),   reflects a change in the
ionization energy due to electronic correlations. 
Indeed as explained {\sl e.g.} by  Almbladh and von Barth\cite{almbladh1985,sm}
the density decay rate, $\kappa$,  is related to the ionization
energy. In particular in the lattice one can show that \cite{sm}
$\kappa=2\cosh^{-1}\lf(I^N/2t\rg)$. 
An accurate computation of the density profile in the
vacuum region is what allow us to compute the absolute value of the
Kohn-Sham potential in the bulk. More precisely the potential in the
bulk is referred to the vacuum site furthest to the interface which is
assumed to have zero Kohn-Sham potential.

The inset of Fig.~\ref{fig-gaphub} shows schematically the behavior
of the Kohn-Sham bands in a large Hubbard chain which can be solved
exactly with periodic boundary conditions\cite{lieb1968}.
The charge is
uniform and thus the Kohn-Sham potential is a constant\cite{notemott}
which, without the vacuum,  remains undetermined. 
  However we know  that the chemical potential as a
function of filling has a jump at half-filling equal to 
the Mott-Hubbard gap $\Delta_{Mott}$. 
If we loosely consider the atoms of the Hubbard chain to have a large constant 
binding energy $v_x=-w_0$ and to be  immersed in a ``vacuum'' with zero
binding energy we expect that the ionization energy
will have a jump at half-filling due to the jump in the
chemical potential. 
Due to DFT Koopman's theorem the center of the bands and 
 Kohn-Sham potential will have a jump at half-filling such
that  $\Delta_{xc}=\Delta_{\rm Mott}$ as shown schematically
in the inset of Fig.~\ref{fig-gaphub}.
In Fig.~\ref{fig-gaphub} we also see that in spite of the bulk chain being short ($L_B=6$), $\D_{xc}$
approximately coincides with $\Delta_{Mott}$ for the infinite system
calculated by Bethe Ansatz showing that this
picture\cite{gunnarsson1986,lima2002}
 is indeed correct and finite size corrections to 
$\Delta_{xc}$ are negligible. 

In Figure \ref{fig-pop} we also note the appearance of a peak at the
boundary between vacuum and bulk, on the vacuum side with width
and height depend on the filling. 
Just as the leading (smallest) decay rate of the wave function is
determined by the first ionization energy, ionization from deeper
states will determine subleading decays rates which are important at short
distances\cite{almbladh1985}.     
Thus to better understand the origin of this peak it is useful to compare the
photoemission spectrum of the Hubbard model and the Kohn-Sham spectrum. 
At large $U$  and for $N=L_B+1$ particles the removal spectra of both
systems is very different as  shown
schematically in the inset of Fig.~\ref{fig-pop}.
In the Hubbard model only two states are available at low
energy\cite{eskes1991} while in the Kohn-Sham spectrum we have $L_B+1$ states
available. This large spectral difference would imply different
subleading decay rates, with a tendency of the Kohn-Sham system to have
a charge density larger than in the interacting system close to the boundary. 
 This tendency is compensated by the appearance of the
peak in the Kohn-Sham potential. Thus the anomalous transfer of spectral
weight in the Hubbard model, which is the hallmark of strong electron
correlation\cite{eskes1991},  reflects in the appearance of the barrier.  



\begin{figure}[!t]
\vspace{1cm}
\begin{center}
\includegraphics[width=0.45\textwidth]{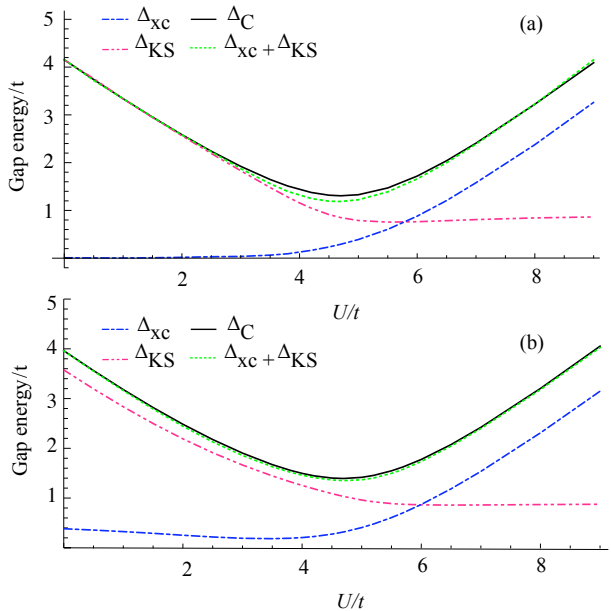}
\end{center}
\caption{Panel (a) and (b)  show  the different contributions to the gap, $\Delta_{KS}$ and  $\Delta_{xc}$ and compare their sum to  
the exact charge gap calculated by Lanczos diagonalization, 
$\Delta_C$. The parameters in the two panels are chosen to have the  same total charge gap at U=0 in the small hopping limit.
Parameters are in panel (a) $\delta=2 t$, $V=0$, in panel (b) $\delta= t$, $V=0.5 t$. In both panels we set $w_0=-6t+U/2$ and 
the potential of the site closer to the bulk-vacuum boundary has been chosen to correct boundary effects. 
} \label{fig-gap}
\end{figure}

Now we consider the transition between a Mott insulator an a ionic
insulator. In order to simulate a binary compound we consider the case
in the presence of 
$\delta$, the  Hubbard $U$ which for simplicity is taken
equal on all atoms and a nearest neighbor repulsion $V$. 
The system shows a transition form a
ionic insulating regime to a Mott insulating regime when $U \sim
2\delta+zV$ with $z=2$ the coordination number\cite{nagaosa1986}. In the atomic 
limit one finds that $\Delta_c^{Mott}=U-2\delta$ in the Mott regime and
$\Delta_c^{Ionic}=2\delta+2zV-U$ in the ionic regime 
with both gaps coinciding at the transition.  Notice that the latter
is larger than the nearest neighbor charge transfer energy 
 corresponding to the excitation of a Frenkel exciton
 $\Delta_{ex}=\Delta_c^{Ionic}-V$ and which becomes relevant bellow. 

Figure~\ref{fig-gap} shows again that that Eq.~\eqref{deltac-app0} is
well satisfied with  negligibly finite size corrections\cite{sm}. 
(a) and (b) show  respectively the  results for
$V=0$ and $\delta=2 t$ and for $V=0.5 t$ and $\delta= t$. As one can easily
check the total charge gap at $U=0$ for small $t$ is the same in the two
cases. However in the first case we
have $\D_{C}\simeq\D_{KS}$ in the ionic insulator and
$\D_{C}\simeq \D_{xc}$ in the Mott-insulating phase, while in the
second case we have a finite contribution of $\D_{xc}$ to the gap in
both regimes. Clearly the appearance of a finite  $\D_{xc}$ in the
ionic regime is 
linked to the presence of the non local interaction $V$. This can be
easily understand by considering the limit of weak tunneling $t<<\Delta_{ex}$.
Using perturbation theory one easily finds that the amount of charge
transferred from odd to even sites is $\delta \rho=
4 t^2/\Delta_{ex}^2$. For a uniform chain, by symmetry, the difference in the 
Kohn Sham potential between even and odd sites is equal to the 
Kohn Sham gap. Applying the same perturbative argument to the
Kohn-Sham system we arrive to the conclusion that 
to match the exact density $\Delta_{KS}=\Delta_{ex}$ therefore  
$\Delta_{xc}=V$.  It is easy to check that these relations are valid in any
dimension. They  are in good agreement with the
numerical results of  Fig.~\ref{fig-gap} in the ionic regime. 

In general we expect that in strongly ionic insulators to a
good approximation the Kohn-Sham
gap matches the first Frenkel exciton and that $\Delta_{xc}$ is given
by its binding energy respect to the fundamental gap. 
 While in ionic salts the Frenkel exciton is
easily accessible experimentally the fundamental gap is difficult to
measure and is often obtain by a theoretical fit to the observed
optical spectra\cite{rohlfing2000}. In any case matching of the Frenkel gap
by $\D_{KS}$ puts a strong constraint on density functionals in strong
ionic insulators. 

To conclude we have computed the exact exchange correlation potential
of a correlated extended system including the (usually undetermined) absolute 
value respect to a vacuum level. This has allowed the first explicit numerical test of
Eq.~\eqref{deltac-app0} in a model ionic/Mott insulator which 
dissipates any possible doubt on the validity of this equation or of the
underling DFT-Koopmans theorem. 
For Mott insulators we have shown that the discontinuity of the
exchange correlation potential is given by the Mott Hubbard gap which
is of the order $U$ for strong correlation. 
On the other hand in a strong ionic
insulator the discontinuity is determined by the nearest neighbor repulsion
$V$ which provides a simple estimate of this elusive quantity.    
In addition we have shown that a surface correlation barrier appears in the
effective potential of a correlated system when the removal spectrum
of the system is very different from the removal spectrum of the
Kohn-Sham system as is expected to occur in electron doped Mott
insulators.


This work was supported  by the Italian Institute of Technology through the project NEWDFESCM. V.B. is indebted to L. Chiodo for discussions.


\begin{thebibliography}{10}
\bibitem{notedeltaxc}
Other authors define $\Delta_{xc}$ from Eq.~\eqref{deltac-app0} which makes it
  trivialy fulfilled. The present definition, instead, preserves the full
  physical significance of the equation.
\end{thebibliography}
\end{document}


\title{Supplementary material to
``Exact exchange-correlation potential of a Ionic/Mott
  insulator with a free surface.''}

\author{Valentina Brosco, Z.-J. Ying, J. Lorenzana}
\affiliation{ISC-CNR and Dipartimento di Fisica, University of Rome 
``La Sapienza'', P.le A. Moro 2, I-00185 Rome, Italy }
\maketitle

\section*{Exact Exchange correlation potential and proof of Koopmans theorem in the lattice}
To calculate the exact Kohn-Sham (KS) potential, {\sl i.e.} the effective non-interacting potential 
 which corresponds to the exact density, we adopt the following strategy:
we first obtain the ground-state density by applying Lanczos diagonalization \cite{ALPS} to the ``bulk+vacuum'' chain, 
we then extract the KS potential by minimizing the difference between the KS
density and the exact one for all values of the KS potential.

The KS density is as usual expressed  in terms of KS orbitals,
$\varphi_i(x, \sigma)$, as  
 $$\rho_{KS}(x)=\sum_{i,\sigma}|\varphi_i(x, \sigma)|^2.$$
 The orbitals $\varphi_i(x, \sigma)$ are in turn defined through the
 well-known KS equations, 
\be\label{KSE}
(\hat T_s+ v_{KS}[x;\rho] )\varphi_i(x, \sigma)=\varepsilon_i\varphi_i(x, \sigma)\ee
where $\hat T_s $ is  defined by 
$\hat T_s\varphi_i(x, \sigma)\equiv -t(\varphi_i(x+1, \sigma)+\varphi_i(x-1, \sigma)-2\varphi_i(x, \sigma))$, $v_{KS}[x;\rho]$ is the effective KS  potential and $\varepsilon_i$ are the KS energies.
To find the exact KS potential, we thus simply minimize relative mean square error on the density  {\sl i.e.} we calculate:
\be
\min_{v_{KS}(x)}\sum_{x,\sigma} \frac{|\rho(x,\sigma)-\rho_{KS}(x,\sigma)|^2}{|\rho(x,\sigma)|^2}
\ee
where $\rho$ denotes the exact density obtained by
Lanczos diagonalization.
After the minimization the relative error on the density is smaller than $10^{-5}$
{\sl i.e.} $|1-\rho_{KS}/\rho| \lesssim 10^{-5}$. 
As we now show, such a high accuracy is necessary to correctly
describe the asymptotic  
decay of the density  in the vacuum and to satisfy  ``Koopmans theorem'' of DFT.
This theorem, which identifies the highest occupied KS eigenvalue
with the exact ionization energy, 
 has been discussed and proved by several authors for standard  DFT in
 the ``continuum'', here we extend the proof to lattice systems 


To prove ``Koopmans theorem'' in the lattice we follow the route
underlined by Almbladh and von Barth\cite{almbladh} and we introduce the
quasiparticle amplitudes, $f_s(x)=\la N-1,s|c_{x\sigma}|N,0\ra$, where
$|N,0\ra$ and  $|N,s\ra$ denote respectively the   ground  and the
$s$-th excited state of the full $N$-electron Hamiltonian. 
\begin{figure}
\begin{center}
\includegraphics[width=0.5\textwidth]{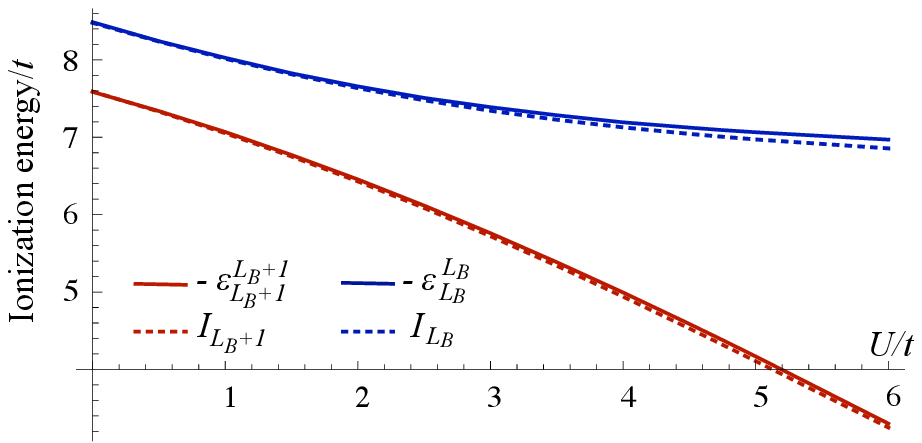}
\end{center}
\vspace{-0.5cm}
\caption{Ionization energy computed from the exact ground state
  energies compared with the highest occupied Kohn-Sham eigenvalue for
  the system with $L_B$ and $L_B+1$ particles  as a function of $U/t$ for constant bulk potential $w_0=-8 t$.   
} \label{ioniz}
 \end{figure}

Considering that in the vacuum sites the density is very small
and decays exponentially with the distance from the surface for our
parameter choice then,  following the same reasoning as in
Ref.~\onlinecite{almbladh}, one can show that many-body effects become
asymptotically irrelevant.  Therefore,  deep in the
vacuum, the quasiparticle  amplitudes satisfy the Schr\"odinger like equation
\be\label{rqpa}
 -t(f^s_{x+1,\sigma}+f^s_{x-1,\sigma}-2f^s_{x,\sigma})-
f^s_{x,\sigma}\Delta E_s =0
\ee
where  $\Delta E_s$  indicates the difference between the ground state
energy of the $N$-particle system and energy $s$-th excited level of
the $N-1$-particle system,  $\Delta E_s=E^N_0-E^{N-1}_s$. 
Solving this recursive equation we obtain the following result for the decay of the  ground-state quasi-particle amplitude, $f_0$ in the vacuum:
\be
f_0(x) \simeq e^{-\beta x}
\ee
with $\beta=\cosh^{-1}(\frac{I_N}{2 t})$, $I_N$ being the ionization
energy $I_N=-\Delta E_0$ of the $N$-particle system. For comparison  in
the continuum the decay rate is given by 
 $f_0(r) \propto e^{-\kappa r} $ with $\kappa=\sqrt{2 m_e I}$  where
 $m_e$ denotes the electron mass.

Now considering that the density can be expressed in terms of the quasi-particle amplitudes as, $\rho_{x\sigma}^{N}=\sum_s f^{*}_s(x\sigma)f_s(x\sigma)$, and that all the quasi-particle amplitudes with $s>0$ decay exponentially faster than
 $f_0$, we can assume that the exponential tail of the density far from the surface of a finite system will be governed by the ground state amplitude $f_0$.
The information on the ionization energy is therefore encoded in the  decay of the density far from the bulk system's surface   
and it can be obtained in  a DFT calculation.
In particular, the relation between the ionization energy and the highest Kohn-Sham eigenvalue becomes 
clear if we consider the analogy between equation \eqref{rqpa} and  Kohn-Sham equations, Eqs. \eqref{KSE}. 
We note indeed that,  since the effective potential can be chosen to vanish in the vacuum far from the surface, 
in  Kohn-Sham formalism the decay of the density is simply governed by the highest occupied Kohn-Sham level which thus concides with the ionization energy. 

The ionization energy as a function of $U$  is shown in
Fig. \ref{ioniz} : we see that Koopmans theorem of DFT is very well
verified, the small error, within 2\%, is due to the finite size of
the vacuum chain.


\section*{Charge gap, Kohn-Sham gap and exchange correlation
  potential discontinuity}
 For sake of completeness we recall here  the derivation of 
the relation between the charge-gap and the exchange-correlation potential shift starting from Koopmans theorem,
this also will allow us to identify the corrections due to finite size effects.

By applying Koopmans theorem  $\Delta_C$ can be rewritten as follows:
\be\label{deltac}
\Delta_C=\Delta_{KS}+\varepsilon_{N+1}^{N+1}-\varepsilon_{N+1}^{N}
\ee
 where  $\varepsilon_{N}^{M}$ 
 indicates the $N$-th Kohn-Sham level of the $M$-particle system  and the gap in the Kohn-Sham spectrum of the $N$-particle system is  given by  $\Delta_{KS}=\varepsilon_{N+1}^{N}-\varepsilon_{N}^{N}$.
%

Using Kohn-Sham equations for the system with $N$ and $N+1$ particles the charge gap can be further partitioned as the sum of four terms, namely
 \be\label{gap-part}
 \Delta_C=\Delta_{KS}+\Delta_{xc}+\Delta_{\varphi}+\Delta_T \ee
 where
$\Delta_{xc}$ is the exchange-correlation gap introduced in the article, essentially equivalent to the one used in the literature (see {\sl e.g.} Ref.[\onlinecite{sham83}])
\be\label{deltaxc}
\Delta_{xc}=\sum_{x}|\varphi^{N+1}_{N+1}(x)|^2\lf(v_{\rm KS}^{N+1}(x)-v_{\rm KS}^{N}(x)\rg)
\ee
while 
$\Delta_{\varphi}$ and $\Delta_T$  are 
due to  the relaxation of the $N+1$-th Kohn-Sham orbital induced by
the addition of a particle, 
\be\label{deltaT}
\Delta_T=\sum_{x}\lf(\varphi_{N+1}^{N+1}(x)\hat T_s\varphi_{N+1}^{N+1}(x)-\varphi_{N+1}^N(x) \hat T_s \varphi_{N+1}^N(x)\rg),
\ee
\be\label{deltafi}
\Delta_{\varphi}=\sum_{x}\lf(|\varphi^{N+1}_{N+1}(x)|^2-|\varphi^{N}_{N+1}(x)|^2\rg)v_{\rm KS}^{N}(x).
\ee
where $v_{\rm KS}^{N}(x)$ denotes the effective potential of the $N$-particle system.
%

For a translationally invariant system
both $\Delta_{\varphi}$ and $\Delta_T$ become of order $1/N$ and can
be neglected in the thermodynamic limit. 
Since  the difference $\delta v_{\rm KS}(x)=v_{\rm
  KS}^{N+1}(x)-v_{\rm KS}^{N}(x)$ becomes constant, in this limit, 
equation \ref{gap-part} reduces to the well-know relation between  the
charge gap and the exchange-correlation potential discontinuity shown
{\sl e.g.} in Ref.~[\onlinecite{perdew83}]. 
%

\begin{figure}[t!]
\begin{center}
\includegraphics[width=0.45\textwidth]{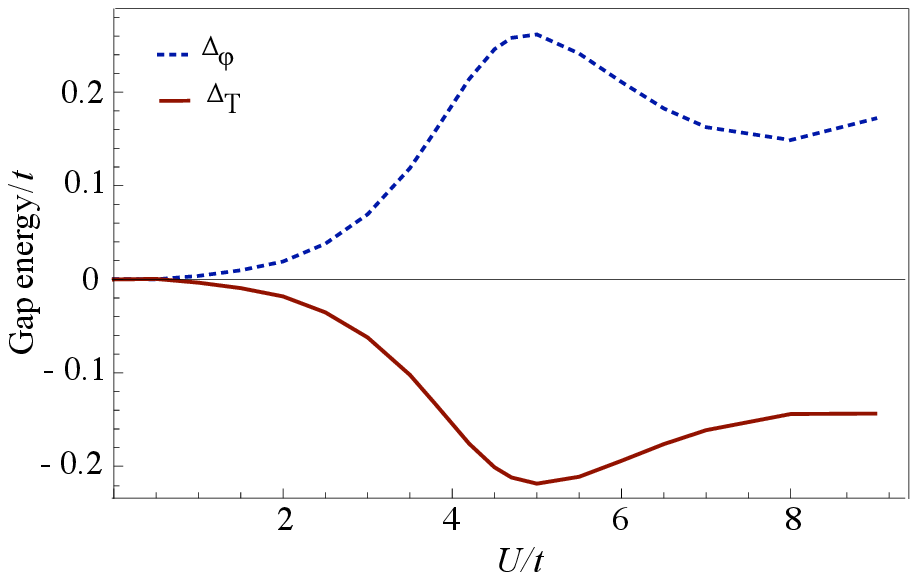}
\end{center}
\caption{Corrections  to the charge gap, $\Delta_\varphi$ and $\Delta_T$ as functions of the ratio, $U/t$.
The bulk potential is $v_x=2 (-1)^x $ while the nearest-neighbor interaction parmeter $V$ equals zero. Similar results are obtained both for $V\neq 0$ and for $v_x=0$.
} \label{gap-corr}
 \end{figure}

 Interestingly we find that  $\Delta_{\varphi}$ and $\Delta_T$ tend to
 cancel in finite systems in all interaction regimes. As an example
 in Figure \ref{gap-corr} we plot the two contributions as a function
 of $U/t$ for the ionic Hubbard model with $v_x=2 (-1)^x $ and $V=0$,
 as one can see the two terms are finite but with opposite signs. 

By applying perturbation theory it is not difficult to show that the
cancellation stems directly from the definition of
exchange-correlation gap given in Eq. \eqref{deltaxc}. 
The effective potential, $v_{\rm KS}^{N+1}$, of the $N+1$ particle system can be indeed related to the potential, $v_{\rm KS}^{N}$, as follows
\be v_{\rm KS}^{N+1}= v_{\rm KS}^{N}+C+\D v \label{pottopot}\ee 
where $C$ is a constant shift, $C=1/(L_B+L_V)\sum_{x}\lf(v_{\rm KS}^{N+1}-v_{\rm KS}^{N}\rg)$, which accounts for the discontinuity and remains finite in the infinite size limit and $\D v$ is a weak site dependent modulation such that $\sum_x  \D v=0$ which vanishes in the infinite size limit. Starting from Eq. \eqref{pottopot}, 
by applying perturbation theory in $ \D v$ we obtain  the following expression for  
$\varepsilon_{N+1}^{N+1}-\varepsilon_{N+1}^{N}$: 
\bea
\varepsilon_{N+1}^{N+1}-\varepsilon_{N+1}^{N}&\simeq& C+\la \varphi^{N}_{N+1}(x)| \D v|\varphi^{N}_{N+1}\ra+\nn\\
& & \!\!\!\!\!\!\!\!\!\!+\sum_{\nu \neq N+1}\frac{|\la \varphi^{N}_{N+1}| \D v|\varphi^{N}_{\nu}\ra|^2}{\ve_{N+1} ^N-\ve_{\nu} ^N}
\eea
Similarly, by inserting Eq. \eqref{pottopot} in Eq. \eqref{deltaxc} we can recast $\D_{xc}$ as 
\be
\D_{xc}=C+\la \varphi^{N+1}_{N+1}| \D v|\varphi^{N+1}_{N+1}\ra \label{dxcsimp}
\ee
Eventually, considering that $\D_{\varphi}+\D_T= \varepsilon_{N+1}^{N+1}-\varepsilon_{N+1}^{N}-\D_{xc}$ 
and expanding the wave-function $|\varphi^{N+1}_{N+1}\ra$ in Eq. \eqref{dxcsimp} we arrive at the following result:
\be
\D_{\varphi}+\D_T\simeq -\sum_{\nu \neq N+1}\frac{|\la \varphi^{N}_{N+1}| \D v|\varphi^{N}_{\nu}\ra|^2}{\ve_{N+1} ^N-\ve_{\nu} ^N}
\ee
From this equation we see that the sum $\D_{\varphi}+\D_T$ is of second order in $\Delta v$, since the change in the wave-function is of first order in $\D v$.

With a similar reasoning one can prove that with the definition of  $\Delta_{xc}$ given in Ref. [\onlinecite{sham83}] the  finite-size errors to second order in $\D v$ have the same modulus but opposite sign.

This work was supported  by the Italian Institute of Technology through the project NEWDFESCM. V.B. is indebted to L. Chiodo for discussions.